\documentclass[aps,prl,twocolumn,groupedaddress]{revtex4}
\usepackage{graphicx}
\usepackage{epstopdf}
\usepackage{url}
\usepackage{color}
\usepackage{dcolumn}
\usepackage{amsmath}
\usepackage{longtable}
\usepackage{subfigure}
\usepackage{dcolumn}
\usepackage{bm}
\begin{document}
\newcommand{\width}{90mm}
\newcommand{\red}{\color{black}}
\newcommand{\redd}{\color{black}}
\title{Data mining and accelerated electronic structure theory as a tool in the search for new functional materials}
\author{C. Ortiz}\thanks{Partly affiliated at Lawrence Berkeley National Laboratory, Berkeley, California during years 2002-2005.}\affiliation{Department of Physics, Uppsala University,  Box 530, SE-751 21 Uppsala, Sweden}

\author{O. Eriksson}\affiliation{Department of Physics, Uppsala University,  Box 530, SE-751 21 Uppsala, Sweden}

\author{M. Klintenberg}\email[Corresponding author: ]{Mattias.Klintenberg@fysik.uu.se}\thanks{Partly affiliated at LBNL during years 1998-2005.}\affiliation{Department of Physics, Uppsala University,  Box 530, SE-751 21 Uppsala, Sweden}
\date{\today}
\maketitle
{\bf
Data mining is a recognized predictive tool in a variety of areas ranging from bioinformatics and drug design to crystal structure prediction. In the present study, an electronic structure implementation has been combined with structural data from the Inorganic Crystal Structure Database to generate results for highly accelerated electronic structure calculations of about 22,000 inorganic compounds. It is shown how data mining algorithms employed on the database can identify new functional materials with desired materials properties, resulting in a prediction of 136 novel materials with potential for use as detector materials for ionizing radiation. The methodology behind the automatized {\it ab-initio} approach is presented, results are tabulated and a version of the complete database is made available at the internet web site {\redd \url{http://gurka.fysik.uu.se/ESP/}} (Ref.\onlinecite{website}).
}

Sensors, solar cells, advanced batteries, and magnetic strips in credit cards are examples of functional materials present in every-day life. One important task for the research in materials science is the continuous improvement and discovery of new such advanced materials. {\redd {\it Ab-initio} electronic structure calculations as a tool for predicting materials properties have steadily increased in use over the years \cite{cohen00} and play today an important role due to the relatively inexpensive and versatile guidance it offers.} There are currently some 8000 studies published annually with this method.

Electronic structure theory applied in materials research is typically done
{\redd in a fashion where a calculation follows, or accompanies, an experimental result}. Knowledge on an atomistic level is thus gained which can help in understanding the experimental results \cite{Olson2000a}. 
In some rarer cases the theoretical calculations predict a materials property which subsequently may be realized experimentally. An example of the latter is the newly proposed tetragonally distorted FeCo alloy with exceptional out of plane magnetic anisotropy \cite{burkert:027203,andersson:037205}.
These materials simulations are done in a one-by-one mode, where one calculation accompanies one experiment. However, with an increasing demand of an accelerated speed in finding or predicting new materials this may not be the most efficient {\redd approach}. Alternatives to this methodology have indeed been discussed, for instance, numerical algorithms which obey evolutionary principles borrowed from biology have been applied to find structural data of compounds and alloys \cite{Hart:2005lr}, 
and in a somewhat similar study where formation probabilities derived from correlations mined for in experimental data were used to guide {\it ab-initio} calculations for unknown structures \cite{ceder}.

In this article a method for automatizing the generation of a new database with electronic structure results for a large number (22,000) of inorganic compounds is presented. The necessary structural information for the {\it ab-initio} calculations is extracted from the Inorganic Crystal Structure Database (ICSD) \cite{icsd}. The electronic structure results are generated within the Local Density Approximation (LDA) of Density Functional Theory (DFT) in combination with a {\redd highly accurate} full potential linear muffin-tin orbital (FP-LMTO) method \cite{wills}.
 
We describe how data mining algorithms can be applied to the database when searching for any particular class of potentially new advanced materials. This may involve semiconductors with tailored band gaps for solar cells and light harvesting, i.e  materials with desired optical properties, and magnetic compounds for energy conversion or magnetocaloric applications.
To demonstrate in detail the power of data mining algorithms with automatic electronic structure calculations we also give a full account of the search and identification of 136 novel compounds with potential use as radiation detector materials. {\redd In addition, we prove a high success rate of the algorithm by an un-biased identification of several known, successful cerium activated materials.} 

Radiation detection systems are generally used in areas such as biomedical imaging, nuclear security, nuclear non-proliferation and treaty verification, and in industry. The limiting factors for the performance of these systems are found within the detector material, and improvements are desired in properties like energy resolution for isotope identification in nuclear security, temporal resolution in biomedical imaging, or simply effective detection in small {\redd sized systems}.

Standard simulations involve to a large extent manual work, where input data files must be generated. We have circumvented this time consuming step by a fully automatic method, where all data files are computer generated after certain materials specific criteria. In addition we have designed a software system which, once all necessary input files have been generated, carries out the simulation automatically {\redd (optimization of parameters, truncation criteria, decision making, etc.). It should also be noted that the algorithm employed has to undertake several steps of learning, in order to decide on a proper set of parameters. For instance, it was found that} the definition {\redd of the base geometry and basis set used in these calculations\cite{wills} needed to be updated after several trial calculations, in order to ensure accuracy of the electronic structure and total energy of the material. }
Hence all steps of the simulations of this work have been made by artificial intelligence and high performance computation. 

The algorithm which executes first principles calculations, with general rules for the computational details as described in Supplementary Information 1, have been applied to some 22,000 compounds from the ICSD database, and the results of these calculations are available on the web site of Ref.\onlinecite{website}. {\red There are a number of electronic structure databases available on the web, e.g. \cite{db1,db2,db3,db4}}, however all at least two orders of magnitude smaller and with different focus. The web-based databases all complement each other.

It should be noted that the crystallographic data from the ICSD originates from different experimental settings, with small variations, giving rise to slightly different electronic structure results {\redd and that several entries in ICSD must be disregarded because at least one site has non-trivial occupancy.} 
The control files for our calculations are available upon request. 

{\redd The crystallographic data needed to construct the control files involves information about the cell geometry, bravais lattice, and the coordinates for each atom and space group. This information is available from the ICSD database \cite{icsd} in, for example, the CIF (crystallographic information file) format. With access to the CIF files the coordinates can then be unfolded and transformed to the minimal bravais lattice\cite{bradleycracknell}. Our approach can make use of any electronic structure method and can be applied to any compilation of structural geometries, even hypothetical ones which have not yet been identified.} 

For each entry in Ref.\onlinecite{website} the electronic structure results are presented as figures illustrating  band structure, density of states (DOS), partial DOS, and charge density contour maps; furthermore, properties like density, {\redd total energy, Fermi energy and band gap (if available) are also listed.} {\red Note that the density is calculated using the experimental lattice parameters. Optimizing the lattice paprameters, i.e. calculating the bulkmodulus, as well as performing spin polarized calculations  will be subjects of future work.}  
 
We now proceed with a detailed example on how mining algorithms on the electronic structure information in Ref.\onlinecite{website} may be used for identifying novel scintillator materials. {\redd The general philosophy of the mining algorithm is to compare specific electronic structure related properties of a larger set of compounds (i.e. the data in Ref.1) to a peer group, which is known to have desired properties connected to a certain functionality of the material.}
We focus here on suitable candidates for nuclear radiation detector materials, and have chosen two sub-groups of materials: (1) cerium activated scintillating materials, and (2) activated semiconductor materials, e.g. Ga doped ZnO (ZnO:Ga)\cite{lehman}. In identifying principles of data mining for these materials, we consider experimental information regarding characteristic electronic properties for known cerium materials, that show $5d$ to $4f$ luminescence, as well as known semiconductor materials which have been found to have encouraging materials properties. The data mining results in 136 candidate materials proposed for further investigation. 

A first desirable property for the materials of interest for this study is high detection probability in small sized units, which is associated to the number of available electrons per unit volume. {\redd A high density and high atomic number (Z) are therefore desired \cite{knoll}. Moreover, a short attenuation length is needed and it is also advantageous that photons scatter mainly through the photoelectric channel. These two properties can be characterized with the photoelectric attenuation length (PAL). PAL is the ratio between the calculated attenuation length ($\lambda=FW/(\rho\cdot[\sigma_{pe}+\sigma_{C}]$) of the incoming radiation in the material and the calculated fraction between the photoelectric ($\sigma_{pe}$) to Compton scattering ($\sigma_{C}$) cross sections (or rather, the ratio $\sigma_{pe}/[\sigma_{pe}+\sigma_{C}])$ at some energy, e.g. 511 keV, which is the energy scale relevant for positron emission tomography (PET)\cite{knoll,atalla}. FW is the formula weight and $\rho$ is the calculated density of the material. The atomic masses and photoelectric- and Compton scattering cross sections are measured, element specific entities and are listed in Supplementary Information 1. PAL summarizes attenuation length and the efficiency of the photoelectric scattering channel and the lower the PAL value is, the higher is the chance that an incoming $\gamma$-ray is absorbed in the material after a short distance by the photoelectric effect, which makes the material more relevant to our study.} 

\begin{figure}
\centering
\includegraphics*[angle=0,scale=0.4]{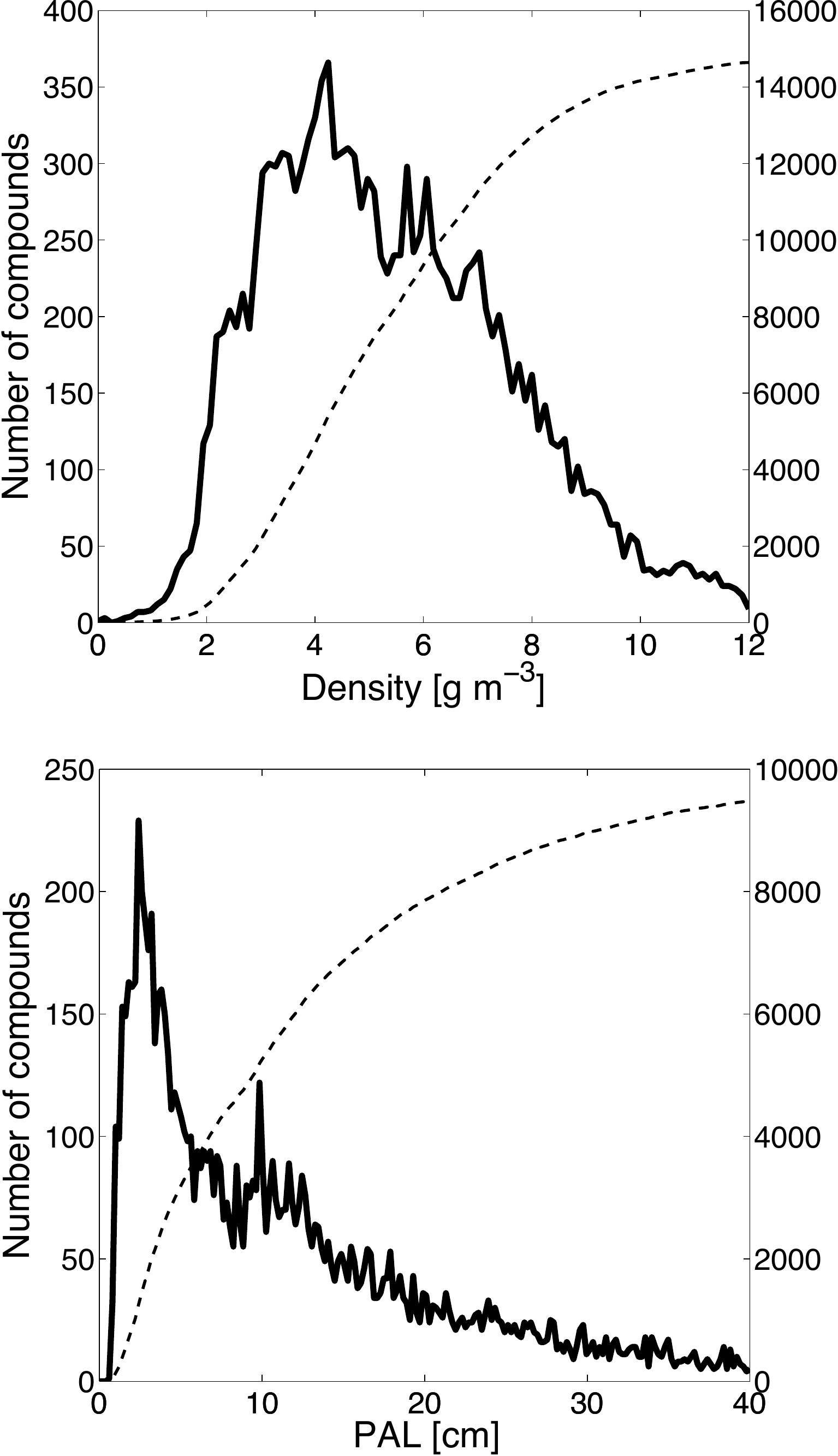}
\caption{Profiles for the density (upper graph) and PAL (lower graph) distributions in energy (full line, scale is to the left of the figures) and their cumulative sum (dashed line, scale is to the right of the figures).}\label{density_PAL_histogram}
\end{figure}

We show in Fig.\ref{density_PAL_histogram} the distribution and cumulative sums of materials densities and the PALs. A high density requirement 
 is imposed as $\rho>6.5\;\mathrm{g/cm^3}$, and we 
 find 4,602 materials satisfying this criterion. As an upper limit for the PAL, the value of a well-known detector material is used, i.e. Tl doped NaI  with \mbox{PAL$_{NaI:Tl}=17\;\mathrm{cm}$}. This limit is satisfied for about 87\% of the materials also satisfying the high density condition and the selection is reduced to 3,983 entries.

\begin{figure}
\centering
\includegraphics[width=60mm]{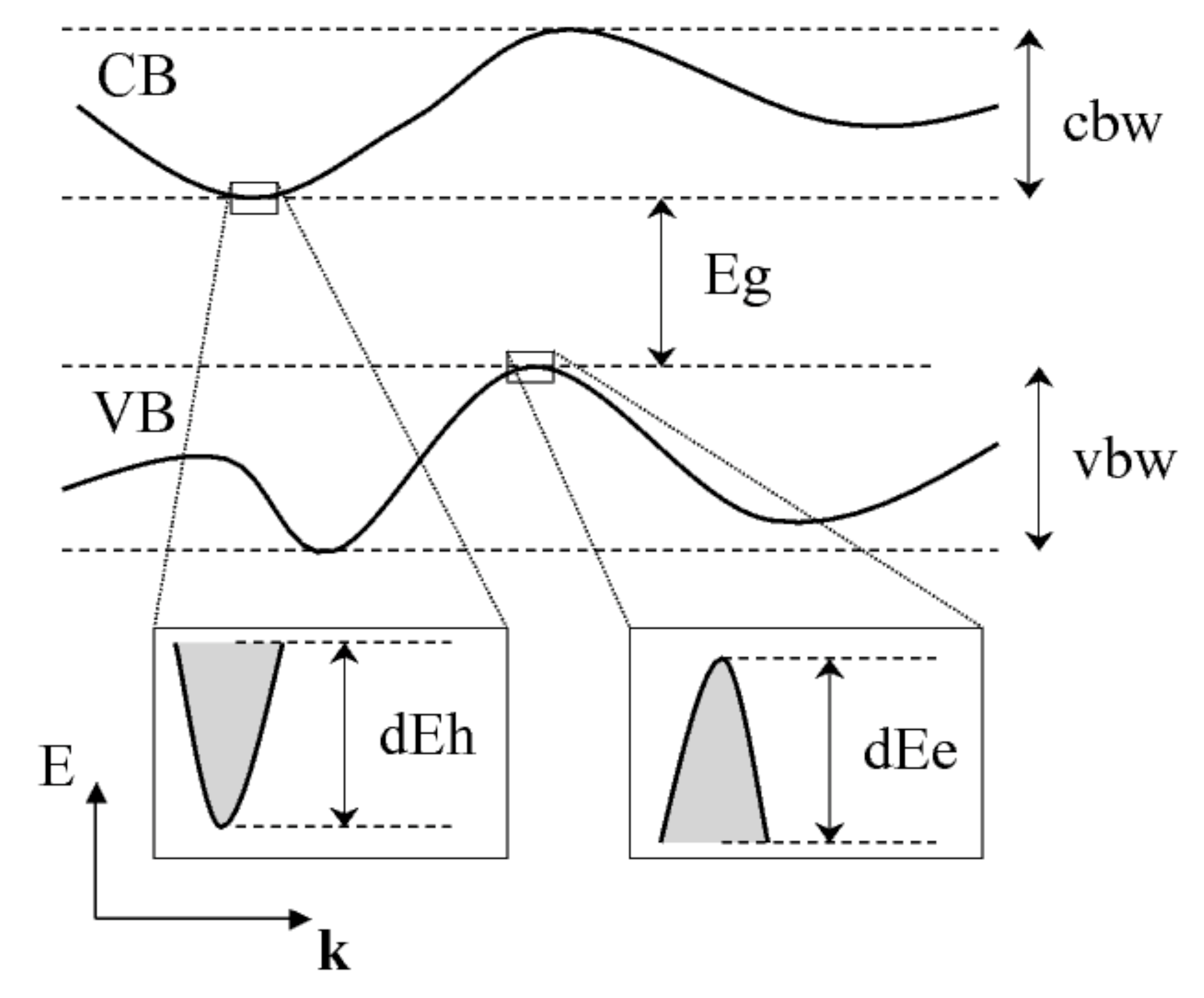}
\caption{Illustration of band widths and  dispersion relations used in the mining algorithms. Because the values for the lowest of CB and the highest of VB occur at different positions in reciprocal space, an indirect band gap (Eg) is shown. By integrating the total density of states around the Fermi level the width of the last occupied electron (dEe) and the width of the first empty state containing one electron (dEh) are found. $vbw$ and $cbw$ represent the width of the valence and conduction band, respectively.}\label{ES_egenskaper}
\end{figure}

We are now left with some four thousand compounds which according to the density and PAL criteria might be suitable detector materials. The synthesis and testing of thousands of candidate compounds is clearly not realistic, {\redd nor is it feasible with standard, manually controlled computational methods to go through such a large} {\redd body of potential materials, to try to identify successful} compounds. {\redd Hence, more efficient methods are needed, where efficient algorithms for calculating the electronic structure must be combined with data mining techniques.}   
{\redd Using a peer group of materials (discussed below) the data mining algorithms can learn selection rules related to electronic structure properties as given by the band structure or the density of states. We will here use the LDA bandgap ($E_g$), the width of the highest valence band ($vbw$), the width of the lowest conduction band ($cbw$), the width of the highest occupying electron in the valence band ($dEe$), and the width of the lowest available state in the conductio band ($dEh$) to further narrow down the list of candidate materials. The definitions of these properties are shown in Fig.\ref{ES_egenskaper}. It should be noted that the parameter 
$vbw$ measures how delocalized} the highest {\redd energy band of the valence states is. In the same way $cbw$ measures the degree of delocalization of the lowest band amongst the conduction band states. The remaining two parameters, $dEe$ and $dEh$, give related information, and stand in direct proportion to the effective mass of the highest electron state and the lowest hole state, respectively. As a matter of fact our mining algorithm could have made use of effective electron and hole masses, instead of $dEe$ and $dEh$, without any change in the result of identified materials.}
When considering the bandgaps a distinction is made between a direct and an in-direct gap material, depending on whether or not the highest energy in the valence band is found at the same point in reciprocal space as the lowest energy in the conduction bands.

The remaining step in the mining algorithm is to make a comparison of a profile of a selected property, {\redd e.g. the valence band- and conduction band-width (vbw and cbw, respectively), for the approximately four thousand compounds in our database which satisfy the density and PAL criteria, with the corresponding profile for a peer group of materials.} This peer group must, of course, have desired properties for the specific functionality desired in the study in question. For semiconducting materials  the peer group is composed of 
well-known materials from Ref.\onlinecite{crchandbook}. 

\begin{figure}
\centering
\includegraphics*[angle=0,scale=0.45]{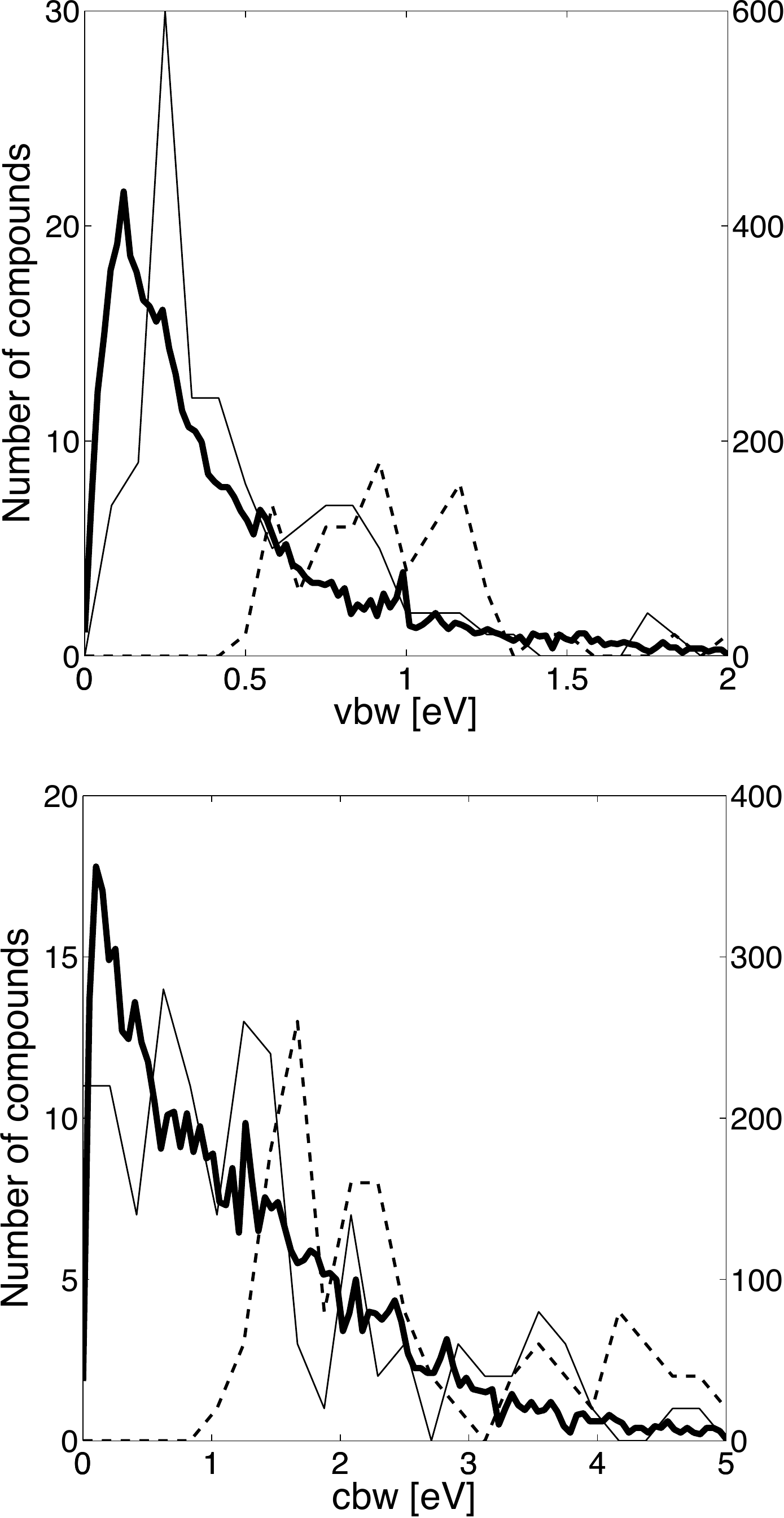}
\caption{Profiles for the $vbw$ (upper graph) and $cbw$ (lower graph) distributions in energy. The thick full line shows the distribution for all compounds in Ref.\onlinecite{website} (the scale is on the right of each plot). The thin full line and the \mbox{dashed} line show the profiles for cerium- and semiconductor materials, respectively, defining the peer groups (the scale is on the left of each plot).}\label{width_histogram}
\end{figure}

\begin{figure}
\centering
\includegraphics*[angle=0,scale=0.45]{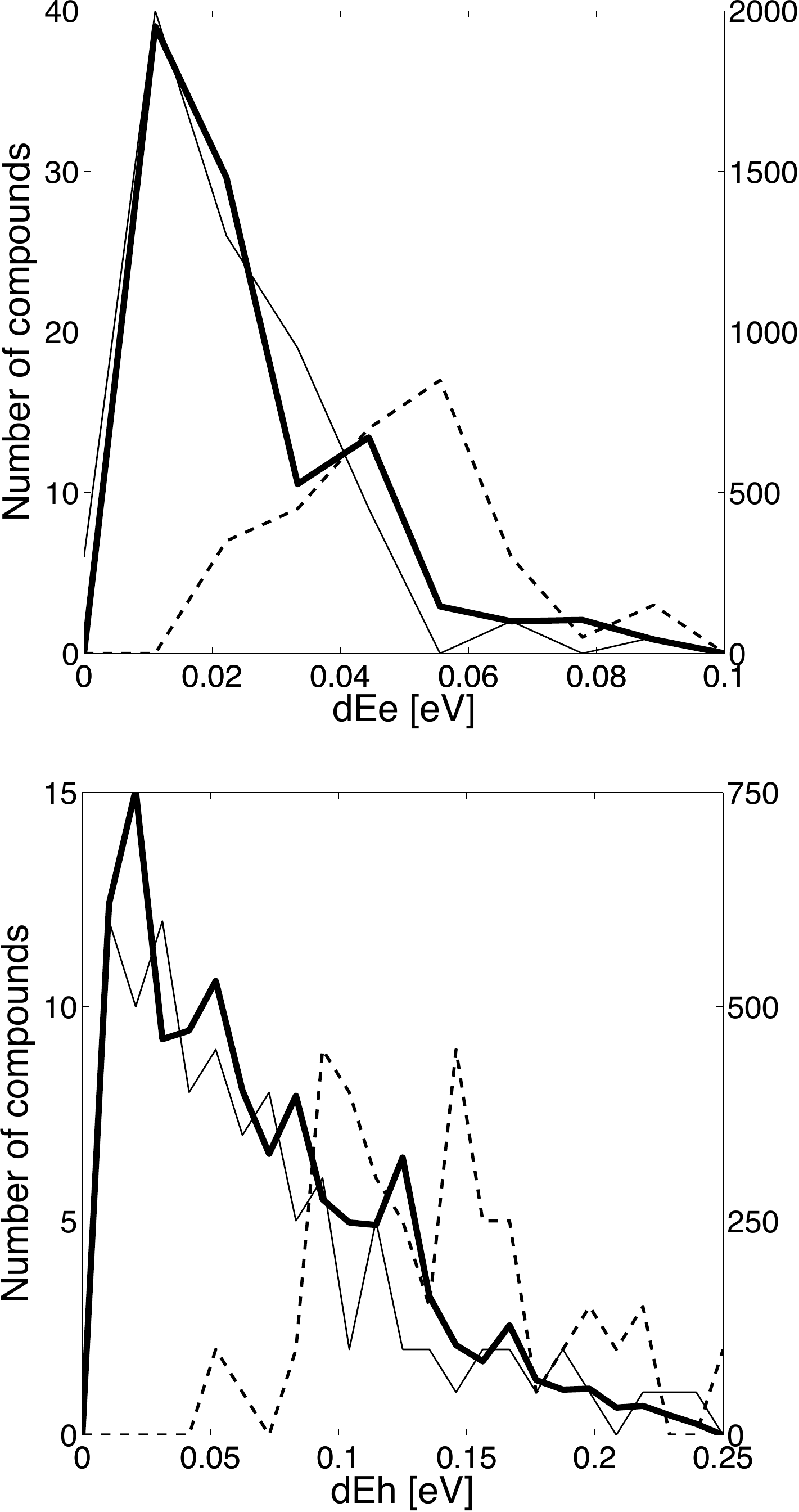}
\caption{Profiles for the $dEe$ (upper graph) and $dEh$ (lower graph) distribution in energy. The thick full line shows the distribution for all compounds in Ref.\onlinecite{website} (the scale is on the right of each plot). The thin full line and the \mbox{dashed} line show the profiles for cerium- and semiconductor materials, respectively, defining the peer groups (the scale is on the left of each plot).}\label{dE_histogram}
\end{figure}

The profiles of $vbw$ and $cbw$ are shown in Fig.\ref{width_histogram}. {\redd In the case of semiconducting materials for use as scintillators, the mining algorithm concludes from the peer group that the $vbw$ value should be greater than \mbox{0.4 eV}. Analogously, $cbw$ must be greater than \mbox{0.9 eV}.
Fig.\ref{dE_histogram} shows the distributions of $dEe$ and the $dEh$, respectively. From the profile of the peer group of semiconductors, the lower limit for the dEe rule is set to \mbox{0.02 eV}. Analogously  the lower limit for the $dEh$ rule is set to \mbox{0.03 eV}. }


For cerium activated scintillating materials a more complex process is known to take place, and for this reason somewhat different mining rules are used. The scintillating process can be thought to occur as follows; a tri-valent cerium atom captures a hole and goes to the tetra-valent state ($Ce^{3+} + h^{+} \rightarrow Ce^{4+}$). A subsequent  capture of an electron takes the tetra-valent cerium atom to an excited tri-valent state ($Ce^{4+} + e^{-} \rightarrow (Ce^{3+})^{*}$). The process can be thought of as an excitation of the Ce $4f$-{\redd electron} into the Ce $5d$-{\redd state}. The band gap of the host material must be large enough to properly accommodate the Ce $4f$ and $5d$ {\redd states} \cite{yen}. Finally, the excited state of the tri-valent Ce atom relaxes to the ground state ($(Ce^{3+})^{*} \rightarrow Ce^{3+} + h\nu$), with the emission of a photon ($h\nu$), ideally around 3 eV, which can be detected with conventional photo-electronics. 
To properly accommodate the Ce $4f$ and $5d$ states a large band gap, \mbox{$>3$ eV}, for the host material is required, and as we will see, this is an important parameter for identifying Ce based scintillator materials. It turns out that the mining rules dEe and dEh are not significant for these materials, and have hence not been used.
For this class of systems our peer group is composed of 
compounds published by Dorenbos in Ref.\onlinecite{dorenbos}.

\begin{figure}
\centering
\includegraphics*[angle=0,scale=0.33]{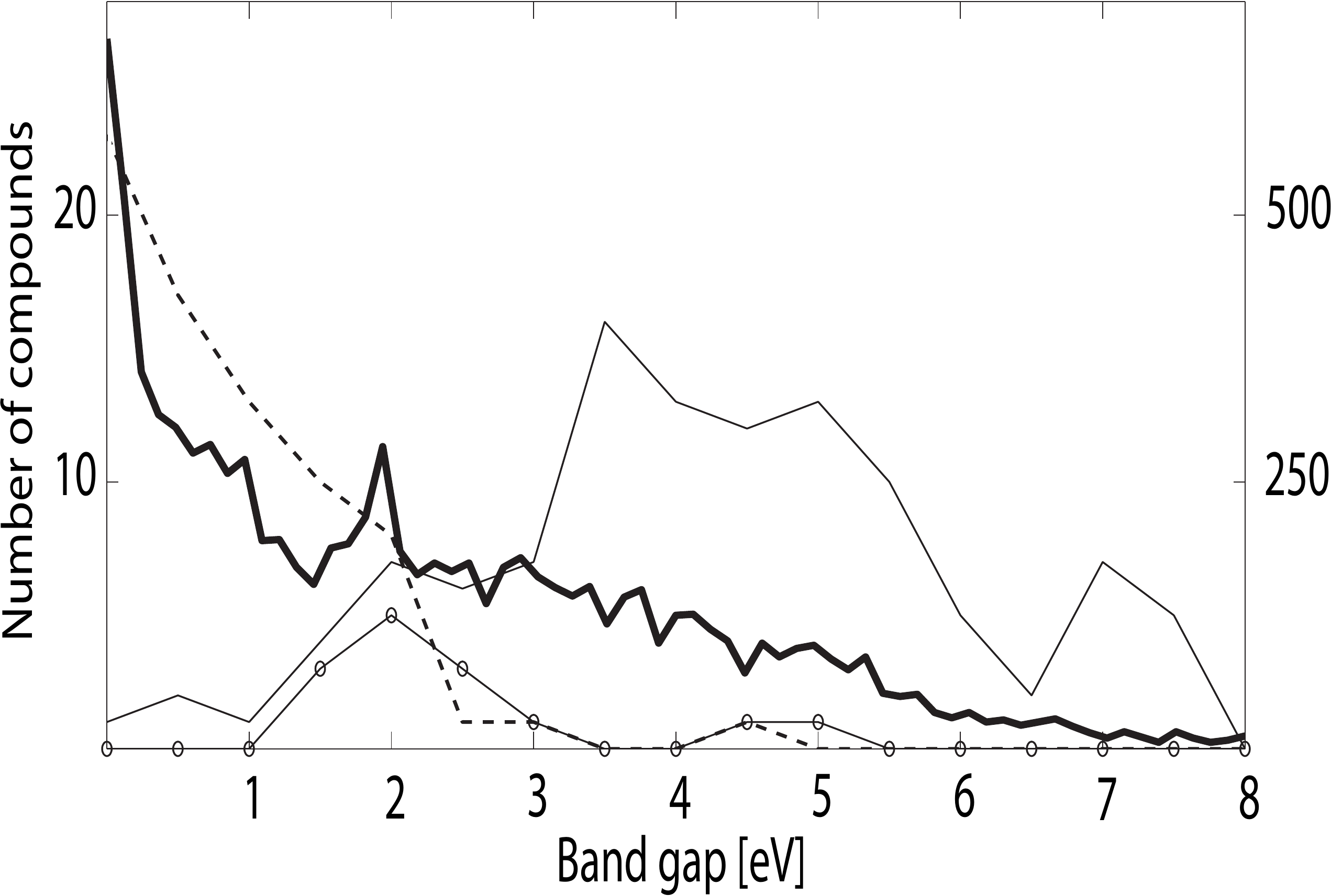}
\caption{Profile for the LDA bandgap distribution for all compounds in Ref.1 (thick full line, the scale is on the right of the figure). Most known cerium activated materials that emit light have an LDA band gap larger than \mbox{3 eV} (thin full line, scale is to the left of the figure). The thin full line with circle markers  show that cerium activated materials where the host contain sulphur can emit light even though the bandgap is small (scale is to the left of the figure). Semiconducting materials from the peer group are shown as a dashed line (scale is to the left of the figure).}\label{LDA_histogram}
\end{figure}

{\redd Before describing the details of the analysis of band gap properties, we note that first principles calculations normally do not reproduce band gaps with great accuracy, whereas the presence of a gap is with very few exceptions always reproduced by such calculations. By now it is well established that the type of theory presented here results in a band gap which is approximately 50 \% of the measured one. Hence it is quite possible to use the calculated band gap as a screening parameter, as long as one makes use of calculated gaps both for the peer} group and {\redd the group of compounds one performs the data mining for.}
Fig.\ref{LDA_histogram} shows the band gap profiles for all compounds with a calculated LDA band gap in Ref.\onlinecite{website} (thick full line), for peer groups of Ce doped materials showing cerium $5d \rightarrow 4f$ emission (thin full line), semiconducting materials (dashed line), and additionally a profile where Ce doped materials which also contain sulfur is shown (thin full line with circle markers). The reason for introducing the latter profile is that 
a detailed analysis of the materials listed in Ref.\onlinecite{dorenbos} shows that host materials containing sulfur allow cerium 5d$\rightarrow$4f emission even though the band gap is smaller compared to Ce materials without sulphur. Therefore the LDA band gap mining rule sets ¨the lower limit of \mbox{3 eV} for Ce activated host compounds, provided that this number is relaxed down to \mbox{1 eV} if the material contains sulfur. 
{\redd In addition a band gap below 4 eV is, as shown in Fig.5, relevant for semiconductor detector materials. For these systems we also imposed a lower limit of the band gap to 0.4 eV, since then thermal noise of the detector is reduced.

The mining algorithm defined for activated semiconducting scintillators is now applied to all materials of Ref.\onlinecite{website} which pass though the PAL and density criterion, and we have been able to identify 66 candidate semiconducting materials. Table 1 in Supplementary Information 2 shows how the number of candidate compounds are reduced as the mining algorithm progresses.} We see that the incorporation of electronic structure related information in the data mining greatly reduces the number of candidate materials. The requirement of a direct band gap with a width in the range of 0.4 - 4 eV, rules out most of the materials in Ref.1, resulting in 195 compounds. 

{\redd Activated semiconductor scintillators represent a very new and novel class of materials and it remains to be seen if these ultra-fast detector materials can be made highly luminous. Among the currently best ultrafast activated semiconductors one finds CuI, ZnO and CdS \cite{lehman}. For these systems the electronic bandgap is direct and wide, and in addition $cbw$ is large. The list of predicted materials in Table 1 of Supplementary Information 3 have similar bandgaps and values of $cbw$ as these known compounds.} {\redd However, it should be noted that in the group of CuI, ZnO and CdS, only CuI is found in Table 1 of Supplementary Information 3, because of the very stringent set of criteria imposed in the data mining procedure (in particular, it is the requirement of a high materials density which} {\redd excludes ZnO and CdS from our list). These criteria are designed to substantially improve the currently used materials in scintillator applications. Hence the predicted list of materials have potential to be substantially better than what is used in current technology. We end the discussion of semiconducting scintillators with a short analysis of the crystal chemistry of the identified materials. First we note that space group 129 (tetragonal) and 62 (orthorhombic) are the most common crystal structures found in Table 1 of Supplementary Information 3, and that materials with space group 141 (tetragonal), 164 (trigonal) and 225 (fcc) are also rather frequent. Secondly, we note that oxides constitute over 50 \% of these materials. We also observe that ionic bonds are present is almost all the materials listed in Table 1 (as well as in Table 2) of Supplementary Information 3. This is clearly a significant result, which is worthwhile to pursue experimentally, and is a unique feature of our study, since it would be difficult to draw this conclusion with any other technique.

Applying the appropriate mining algorithm for Ce activated compounds to all materials of Ref.\onlinecite{website} identifies some 70 candidate materials. We note again that the rules learned by the mining algorithm are optimized to identify new materials with superior properties compared to known systems. Table 2 in Supplementary Information 2 shows how the number of candidate compounds is reduced as the data mining progresses, and it is clear that the band gap rule stands out as the most efficient. The resulting list of compounds identified as potential hosts for Ce (Ce activated scintillator materials) are given in Table 2 of Supplementary Information 3. For cerium activation, compounds containing Gd, Lu, Y, La and Sc are known to often be efficient scintillators, the reason being that these elements do not introduce electron/hole traps and therefore allow efficient energy transfer to the cerium dopant. Hence, it it is gratifying that among the 
materials in Table 2 of Supplementary Information 3, many are indeed rare-earth (RE) based. The quantum efficiency of the cerium $5d$ to $4f$ transition is largely dictated by the chemical bonding between the cerium and host atoms, a property which is ideal to investigate with electronic structure methods. The fact that the crystal structure as well as type of host cation are important parameters is reflected in that most useful cerium activated compounds today are orthosilicates, aluminates, phosphates or simple metal-halides. 
We note that several orthosilicates, aluminates, phosphates or simple metal-halides are predicted in
Table 2 of Supplementary Information 3.

As regards crystal structure we note that space group 62 constitutes a large group of Ce doped compounds (this space group was also prevalent for the semiconducting scintillators). Space group 14 (monoclinic) and 225 (fcc) are also common in our Table of Ce doped scintillators. Again, and as noted above, the presence of ionic} compounds is {\redd clear from this table.

It might be argued that a theoretical prediction of novel materials with improved properties is uncertain, if it is not accompanied by an experimental verification. To work around this argument without doing any real synthesis and measurements it is useful to consider the following. Suppose some of the most successful/useful cerium activated detector materials, {\it e.g.} LaBr$_{3}$, LaCl$_{3}$, LaF$_{3}$, Lu$_{2}$SiO$_{5}$, Gd$_{2}$SiO$_{5}$, LuPO$_{4}$,YAlO$_{3}$ and CeF$_{3}$,  had never been discovered, would the present method be capable of identifying them? To answer this question we removed the materials listed above from the peer group, thus forcing the mining algorithm to learn from a smaller peer group, to see if our algorithms would identify these well known compounds. The results of this analysis is indeed very encouraging. Four out of the eight materials are immediately identified (LaF$_{3}$, Lu$_{2}$SiO$_{5}$, Gd$_{2}$SiO$_{5}$ and LuPO$_{4}$). LaBr$_{3}$ and Ce
 F$_{3}$ do not appear in our list, because they have too low density and the compounds LaCl$_{3}$ and YAlO$_{3}$ are also excluded since they have too high PAL value. Should the density cut-off be set to 5.0, LaBr$_{3}$ and CeF$_{3}$ would also have been identified as well as several more interesting compounds, {\it e.g.} Ce doped Y$_{2}$Si$_{2}$O$_{7}$. If we had used a higher value of the PAL, in the screening process, we would also have included LaCl$_{3}$ and YAlO$_{3}$ (as well as several other compounds), but it should be noted that the too high PAL value of these two copmpounds is known to make them less attractive as scintilator materials, even though they have possitive features like low cost and are easily synthesized. The exercise described above shows that our mining algorithm and electronic structure method has the desired accuracy for identifying novel materials with desired properties.

Inspection of Table 2 in Supplementary Information 3 reveal several compounds of special interest and AsLuO$_{4}$ and ClGdO stand out in this group, especially beacuse the first is isoelectronic to LuPO$_{4}$ and the latter one is related to the compound BrGdO, which is a known highly luminous phosphor. In fact lanthanide oxyhalides doped with cerium are interesting because also the La and Lu versions are well-known luminous phosphor materials. We note that the successful materials discussed in the previous paragraph all have large LDA bandgaps and this fact indicates that the following materials also deserve special attention: AlO$_{3}$Tb, Al$_{2}$Gd$_{2}$O$_{7}$Sr, Ba$_{4}$O$_{10}$Ru$_{3}$ and O$_{4}$SrYb$_{2}$.

The materials predicted here are the sole result of theoretical modeling and are found by using a data mining algorithm which uses material properties of a peer group of already well-known materials. Obviously the method presented here can be employed to identify materials with other properties, for instance novel materials for fuel cell and battery applications, super hard compounds and magnetic nano-devices with taylormade transport properties.}

\begin{acknowledgments}
We thank Dr. B. Sanyal, Dr. D. \AA berg, and Dr. B. Sadigh for helpful discussions on electronic structure theory, and Dr. M. J. Weber, Dr. S. E. Derenzo, and Dr. W. W. Moses for helpful discussions on radiation detection. This work has been sponsored in part by NNSA/na22; HSARPA; Stiftelsen f\"or internationalisering av h\"ogre utbildning och forskning (STINT); Vetenskapsr\aa det (VR); Kungliga vetenskapsakademin (KVA); SNIC/SNAC and the G\"oran Gustafsson Stiftelse.
\end{acknowledgments}


\begin{widetext}

\section{Appendix A: Suplementary information 1}

Any {\it ab-initio} method requires initial input data for the atomic species and their relative position in the crystal as well as information about truncation in expansion of wavefunctions, density and potential. The structural data are in this work extracted from the ICSD [8].  Additionally for the FP-LMTO method used here [9] we need to define: 
\begin{itemize} 
\item A muffin-tin radius, $R_{MT}$, optimized to be the largest value for non-overlapping neighboring spheres. The initial value for $R_{MT}$ is set to be the ionic radius. The electronic structure calculation is iterated three times and between each iteration $R_{MT}$ is set to the smallest value for where the potential between each atom pair reaches a maximum. 
\item An upper limit, $l_{cut}$, for the expansion of the angular part of the wavefunctions inside the muffin-tin, which we set equal to the highest populated orbital in the valence, plus one (e.g. for sp-bonded materials we use s,p and d orbitals as basis functions). 
\item The expansion of density and potential inside the muffin-tin radius is done up to $l_{cut}$=8. 
\item The grid for the sampling of the irreducible Brillouin zone and the Fourier mesh for expanding the density and potential in the interstitial are all set inversely proportional to the lengths of the crystal axes and to include all high symmetry points. Hence for smaller cells we make use of a higher number of k-points whereas for larger cells we use a smaller number of k-points. 
\item For each atom a selection of which electronic states should be categorized as chemically inert core states, and which are considered to be chemically active valence orbitals must be made. We have here made a conventional choice which is listed in the table below. 
\item In the table below we also list for each atom the experimental crossections for compton scattering ($\sigma_{C}$) and the photoelectric effect ($\sigma_{pe}$), which are used for the calclation of the PAL, defined in the main text.
\end{itemize}

\begin{center}

\begin{longtable}{lllllll}
\caption{Definition of atomic configurations including atomic number (Z), symbol (atom),  core- and valence electron configuration. The atomic masses as well as photoelectric- and Compton crossections are also listed.}\label{tab:material}\\

   \multicolumn{1}{l}{\textbf{Z}} &
   \multicolumn{1}{l}{\textbf{Atom}} &
   \multicolumn{1}{l}{\textbf{core}} &
   \multicolumn{1}{l}{\textbf{Valence}} &
   \multicolumn{1}{l}{\textbf{Atomic mass}} &
   \multicolumn{1}{l}{\textbf{$\sigma_{pe}$}} &
   \multicolumn{1}{l}{\textbf{$\sigma_{C}$}} \\[0.5ex] 
   \\[-1.8ex]
\endfirsthead

\multicolumn{7}{c}{{\tablename} \thetable{} -- Continued} \\[0.5ex]
   \multicolumn{1}{l}{\textbf{Z}} &
   \multicolumn{1}{l}{\textbf{Atom}} &
   \multicolumn{1}{l}{\textbf{core}} &
   \multicolumn{1}{l}{\textbf{Valence}} &
   \multicolumn{1}{l}{\textbf{Atomic mass}} &
   \multicolumn{1}{l}{\textbf{$\rho_{pe}$}} &
   \multicolumn{1}{l}{\textbf{$\rho_C$}} \\[0.5ex] 
  \\[-1.8ex]
\endhead

  \multicolumn{7}{l}{{Continued on Next Page\ldots}} \\
\endfoot

\endlastfoot

  1 &  H & & 1$s^{1}$ &  1.01 & 8.79e-09 & 0.29 \\
  2 & He & & 1$s^{2}$ &  4.00 & 2.70e-07 & 0.57 \\
  3 & Li & [He] & 2$s^{1}$  &  6.94 & 2.62e-06 & 0.86\\
  4 & Be & [He] & 2$s^{2}$  &  9.01 & 1.33e-05 & 1.15\\
  5 &  B & [He] & 2$s^{2}$ 2$p^{1}$ & 10.81 & 4.59e-05 & 1.43 \\
  6 &  C & [He] & 2$s^{2}$ 2$p^{2}$ & 12.01 & 1.22e-04 & 1.72\\
  7 &  N & [He] & 2$s^{2}$ 2$p^{3}$ & 14.01 & 2.75e-04 & 2.00 \\
  8 &  O & [He] & 2$s^{2}$ 2$p^{4}$ & 16.00 & 5.51e-04 & 2.29\\
  9 &  F & [He] & 2$s^{2}$ 2$p^{5}$ & 19.00 & 9.85e-04 & 2.58\\
 10 & Ne & [He] & 2$s^{2}$ 2$p^{6}$ & 20.18 & 1.69e-03 & 2.86\\
 11 & Na & [Ne] & 3$s^{1}$  & 22.99 & 2.62e-03 & 3.15\\
 12 & Mg & [Ne] & 3$s^{2}$  & 24.30 & 3.97e-03 & 3.43\\
 13 & Al & [Ne] & 3$s^{2}$ 3$p^{1}$  & 26.98 & 6.10e-03 & 3.72\\
 14 & Si & [Ne] & 3$s^{2}$ 3$p^{2}$ & 28.09 & 8.46e-03 & 4.00 \\
 15 &  P & [Ne] & 3$s^{2}$ 3$p^{3}$  & 30.97 & 1.18e-02 & 4.29\\
 16 &  S & [Ne] & 3$s^{2}$ 3$p^{4}$  & 32.07 & 1.69e-02 & 4.57\\
 17 & Cl & [Ne] & 3$s^{2}$ 3$p^{5}$  & 35.45 & 2.19e-02 & 4.86\\
 18 & Ar & [Ne] & 3$s^{2}$ 3$p^{6}$  & 39.95 & 2.87e-02 & 5.14\\
 19 &  K & [Ne] & 3$s^{2}$ 3$p^{6}$ 4$s^{1}$  & 39.10 & 3.68e-02 & 5.43\\
 20 & Ca & [Ne] & 3$s^{2}$ 3$p^{6}$ 4$s^{2}$  & 40.08 & 4.91e-02 & 5.71\\
 21 & Sc & [Ne] & 3$s^{2}$ 3$p^{6}$ 3$d^{1}$ 4$s^{2}$  & 44.96 & 5.94e-02 & 5.99\\
 22 & Ti & [Ne] & 3$s^{2}$ 3$p^{6}$ 3$d^{2}$ 4$s^{2}$ & 47.88 & 7.74e-02 & 6.28 \\
 23 &  V & [Ne] & 3$s^{2}$ 3$p^{6}$ 3$d^{3}$ 4$s^{2}$ & 50.94 & 9.47e-02 & 6.56 \\
 24 & Cr & [Ne] & 3$s^{2}$ 3$p^{6}$ 3$d^{5}$ 4$s^{1}$  & 52.00 & 1.16e-01 & 6.84\\
 25 & Mn & [Ne] & 3$s^{2}$ 3$p^{6}$ 3$d^{5}$ 4$s^{2}$  & 54.94 & 1.40e-01 & 7.13\\
 26 & Fe & [Ne] & 3$s^{2}$ 3$p^{6}$ 3$d^{6}$ 4$s^{2}$  & 55.85 & 1.59e-01 & 7.41\\
 27 & Co & [Ne] & 3$s^{2}$ 3$p^{6}$ 3$d^{7}$ 4$s^{2}$  & 58.93 & 1.94e-01 & 7.69\\
 28 & Ni & [Ar] & 3$d^{8}$ 4$s^{2}$  & 58.69 & 2.21e-01 & 7.98\\
 29 & Cu & [Ar] & 3$d^{10}$ 4$s^{1}$  & 63.55 & 2.65e-01 & 8.26\\
 30 & Zn & [Ar] & 3$d^{10}$ 4$s^{2}$ & 65.39 & 3.10e-01 & 8.54 \\
 31 & Ga & [Ar] & 3$d^{10}$ 4$s^{2}$ 4$p^{1}$ & 69.72 & 3.56e-01 & 8.82 \\
 32 & Ge & [Ar] & 3$d^{10}$ 4$s^{2}$ 4$p^{2}$  & 72.61 & 4.14e-01 & 9.11\\
 33 & As & [Ar] & 3$d^{10}$ 4$s^{2}$ 4$p^{3}$   & 74.92 & 4.68e-01 & 9.39\\
 34 & Se & [Ar] & 3$d^{10}$ 4$s^{2}$ 4$p^{4}$  & 78.96 & 5.49e-01 & 9.67\\
 35 & Br & [Ar] & 3$d^{10}$ 4$s^{2}$ 4$p^{5}$  & 79.90 & 6.24e-01 & 9.95\\
 36 & Kr & [Ar] & 3$d^{10}$ 4$s^{2}$ 4$p^{6}$ & 83.80 & 7.09e-01 & 10.24 \\
 37 & Rb & [Ar] 3$d^{10}$ & 4$s^{2}$ 4$p^{6}$ 5$s^{1}$ & 85.47 & 7.98e-01 & 10.52 \\
 38 & Sr & [Ar] 3$d^{10}$ & 4$s^{2}$ 4$p^{6}$ 5$s^{2}$ & 87.62 & 9.16e-01 & 10.79 \\
 39 &  Y & [Ar] 3$d^{10}$ & 4$s^{2}$ 4$p^{6}$ 4$d^{1}$ 5$s^{2}$  & 88.91 & 1.02e+00 & 11.08\\
 40 & Zr & [Ar] 3$d^{10}$ & 4$s^{2}$ 4$p^{6}$ 4$d^{2}$ 5$s^{2}$   & 91.22 & 1.10e+00 & 11.36\\
 41 & Nb & [Ar] 3$d^{10}$ & 4$s^{2}$ 4$p^{6}$ 4$d^{4}$ 5$s^{1}$  & 92.91 & 1.27e+00 & 11.64 \\
 42 & Mo & [Ar] 3$d^{10}$ & 4$s^{2}$ 4$p^{6}$ 4$d^{5}$ 5$s^{1}$ & 95.94 & 1.49e+00 & 11.92 \\
 43 & Tc & [Ar] 3$d^{10}$ & 4$s^{2}$ 4$p^{6}$ 4$d^{6}$ 5$s^{1}$  & 98.00 & 1.57e+00 & 12.20\\
 44 & Ru & [Ar] 3$d^{10}$ & 4$s^{2}$ 4$p^{6}$ 4$d^{7}$ 5$s^{1}$ & 101.07 & 1.74e+00 & 12.48 \\
 45 & Rh & [Kr] & 4$d^{8}$ 5$s^{1}$   & 102.91 & 1.92e+00 & 12.76\\
 46 & Pd & [Kr] & 4$d^{10}$  & 106.42 & 2.15e+00 & 13.04\\
 47 & Ag & [Kr] & 4$d^{10}$ 5$s^{1}$  & 107.87 & 2.36e+00 & 13.32\\
 48 & Cd & [Kr] & 4$d^{10}$ 5$s^{2}$  & 112.41 & 2.60e+00 & 13.60\\
 49 & In & [Kr] & 4$d^{10}$ 5$s^{2}$ 5$p^{1}$  & 114.82 & 2.87e+00 & 13.88\\
 50 & Sn & [Kr] & 4$d^{10}$ 5$s^{2}$ 5$p^{2}$  & 118.71 & 3.06e+00 & 14.16\\
 51 & Sb & [Kr] & 4$d^{10}$ 5$s^{2}$ 5$p^{3}$  & 121.75 & 3.30e+00 & 14.44\\
 52 & Te & [Kr] & 4$d^{10}$ 5$s^{2}$ 5$p^{4}$  & 127.60 & 3.66e+00 & 14.72\\
 53 &  I & [Kr] & 4$d^{10}$ 5$s^{2}$ 5$p^{5}$ & 126.90 & 4.06e+00 & 15.00\\
 54 & Xe & [Kr] & 4$d^{10}$ 5$s^{2}$ 5$p^{6}$  & 131.29 & 4.31e+00 & 15.27\\
 55 & Cs & [Kr] 4$d^{10}$ & 5$s^{2}$ 5$p^{6}$ 6$s^{1}$  & 132.91 & 4.68e+00 & 15.55\\
 56 & Ba & [Kr] 4$d^{10}$ & 5$s^{2}$ 5$p^{6}$ 6$s^{2}$  & 137.33 & 5.03e+00 & 15.83\\
 57 & La & [Kr] 4$d^{10}$ & 5$s^{2}$ 5$p^{6}$ 5$d^{1}$ 6$s^{2}$  & 138.91 & 5.47e+00 & 16.11\\
 58 & Ce & [Kr] 4$d^{10}$ 4$f^{2}$ & 5$s^{2}$ 5$p^{6}$ 6$s^{2}$  & 140.12 & 5.86e+00 & 16.38\\
 59 & Pr & [Kr] 4$d^{10}$ 4$f^{3}$ & 5$s^{2}$ 5$p^{6}$ 6$s^{2}$  & 140.91 & 6.41e+00 & 16.66\\
 60 & Nd & [Kr] 4$d^{10}$ 4$f^{4}$ & 5$s^{2}$ 5$p^{6}$ 6$s^{2}$  & 144.24 & 6.95e+00 & 16.94\\
 61 & Pm & [Kr] 4$d^{10}$ 4$f^{5}$ & 5$s^{2}$ 5$p^{6}$ 6$s^{2}$  & 145.00 & 7.43e+00 & 17.22\\
 62 & Sm & [Kr] 4$d^{10}$ 4$f^{6}$ & 5$s^{2}$ 5$p^{6}$ 6$s^{2}$  & 150.36 & 7.96e+00 & 17.50\\
 63 & Eu & [Kr] 4$d^{10}$ 4$f^{7}$ & 5$s^{2}$ 5$p^{6}$ 6$s^{2}$  & 151.97 & 8.56e+00 & 17.77\\
 64 & Gd & [Kr] 4$d^{10}$ 4$f^{7}$ & 5$s^{2}$ 5$p^{6}$ 5$d^{1}$ 6$s^{2}$  & 157.25 & 9.02e+00 & 18.05\\
 65 & Tb & [Kr] 4$d^{10}$ 4$f^{9}$ & 5$s^{2}$ 5$p^{6}$ 6$s^{2}$  & 158.93 & 9.78e+00 & 18.33\\
 66 & Dy & [Kr] 4$d^{10}$ 4$f^{10}$ & 5$s^{2}$ 5$p^{6}$ 6$s^{2}$  & 162.50 & 1.05e+01 & 18.60\\
 67 & Ho & [Kr] 4$d^{10}$ 4$f^{11}$ & 5$s^{2}$ 5$p^{6}$ 6$s^{2}$  & 164.93 & 1.11e+01 & 18.88\\
 68 & Er & [Kr] 4$d^{10}$ 4$f^{12}$ & 5$s^{2}$ 5$p^{6}$ 6$s^{2}$  & 167.26 & 1.19e+01 & 19.16\\
 69 & Tm & [Kr] 4$d^{10}$ 4$f^{13}$ & 5$s^{2}$ 5$p^{6}$ 6$s^{2}$  & 168.93 & 1.27e+01 & 19.44\\
 70 & Yb & [Kr] 4$d^{10}$ 4$f^{14}$ & 5$s^{2}$ 5$p^{6}$ 6$s^{2}$  & 173.04 & 1.37e+01 & 19.71\\
 71 & Lu & [Kr] 4$d^{10}$ 4$f^{14}$ & 5$s^{2}$ 5$p^{6}$ 5$d^{1}$ 6$s^{2}$& 174.97 & 1.44e+01 & 19.98 \\
 72 & Hf & [Xe] 4$f^{14}$ & 5$d^{2}$ 6$s^{2}$  & 178.49 & 1.48e+01 & 20.26\\
 73 & Ta & [Xe] 4$f^{14}$ & 5$d^{3}$ 6$s^{2}$  & 180.95 & 1.57e+01 & 20.54\\
 74 &  W & [Xe] 4$f^{14}$ & 5$d^{4}$ 6$s^{2}$  & 183.85 & 1.65e+01 & 20.81\\
 75 & Re & [Xe] 4$f^{14}$ & 5$d^{5}$ 6$s^{2}$ & 186.21 & 1.78e+01 & 21.08 \\
 76 & Os & [Xe] 4$f^{14}$ & 5$d^{6}$ 6$s^{2}$  & 190.20 & 1.94e+01 & 21.36\\
 77 & Ir & [Xe] 4$f^{14}$ & 5$d^{7}$ 6$s^{2}$  & 192.22 & 2.06e+01 & 21.64\\
 78 & Pt & [Xe] 4$f^{14}$ & 5$d^{9}$ 6$s^{1}$   & 195.08 & 2.11e+01 & 21.91\\
 79 & Au & [Xe] 4$f^{14}$ & 5$d^{10}$ 6$s^{1}$  & 196.97 & 2.23e+01 & 22.19\\
 80 & Hg & [Xe] 4$f^{14}$ & 5$d^{10}$ 6$s^{2}$  & 200.59 & 2.36e+01 & 22.46\\
 81 & Tl & [Xe] 4$f^{14}$ & 5$d^{10}$ 6$s^{2}$ 6$p^{1}$  & 204.38 & 2.56e+01 & 22.73\\
 82 & Pb & [Xe] 4$f^{14}$ & 5$d^{10}$ 6$s^{2}$ 6$p^{2}$  & 207.20 & 2.65e+01 & 23.01\\
 83 & Bi & [Xe] 4$f^{14}$ & 5$d^{10}$ 6$s^{2}$ 6$p^{3}$  & 208.98 & 2.80e+01 & 23.28\\
 84 & Po & [Xe] 4$f^{14}$ & 5$d^{10}$ 6$s^{2}$ 6$p^{4}$  & 209.00 & 3.00e+01 & 23.55\\
 85 & At & [Xe] 4$f^{14}$ & 5$d^{10}$ 6$s^{2}$ 6$p^{5}$  & 210.00 & 3.15e+01 & 23.82\\
 86 & Rn & [Xe] 4$f^{14}$ & 5$d^{10}$ 6$s^{2}$ 6$p^{6}$  & 222.00 & 3.33e+01 & 24.10\\
 87 & Fr & [Xe] 4$f^{14}$ 5$d^{10}$ & 6$s^{2}$ 6$p^{6}$ 7$s^{1}$  & 223.00 & 3.47e+01 & 24.38\\
 88 & Ra & [Xe] 4$f^{14}$ 5$d^{10}$ & 6$s^{2}$ 6$p^{6}$ 7$s^{2}$  & 226.03 & 3.65e+01 & 24.64\\
 89 & Ac & [Xe] 4$f^{14}$ 5$d^{10}$ & 6$s^{2}$ 6$p^{6}$ 6$d^{1}$ 7$s^{2}$  & 227.03 & 3.82e+01 & 24.92\\

\end{longtable}
\end{center}


\newpage
\begin{table}
\section{Appendix B: Suplementary information 2}
\begin{center}
  \begin{tabular}{llr}
  \hline
  Parameter & Limit & No. remaining \\
  \hline
           & Ref.1                   & 22,283 \\
  Density\footnotemark[1]  & $>$ 6.5 g/cm$^{3}$  & 4,602 \\
  PAL      & $<$ 17 cm             & 3,983 \\
  Gap type & direct                & 334 \\
  Band gap & 0.4 $<$ E$_g$ $<$ 4eV & 195   \\ 
  vbw      & $>$ 0.4 eV            & 121   \\
  cbw      & $>$ 0.6 eV            & 104    \\
  dEe      & $>$ 0.01 eV           & 68    \\
  dEh\footnotemark[2]      & $>$ 0.03 eV           &   66  \\
  \hline
\footnotetext[1]{An upper limit of 13.0 g/cm$^{3}$ is applied for the density.} 
\footnotetext[2]{21 out of the 104 compounds lack values for dEe and dEh which means that only 2 and 12 compounds are removed by these two constraints, respectively.}
\end{tabular}
\caption{Results of the mining algorithm for wide-gap semiconductor materials. A final list of 66 compounds is obtained.}\label{tab:semimining}
\vspace{1cm}
  \begin{tabular}{llr}
  \hline
  Parameter & Limit & No. remaining \\
  \hline
           & Ref.1                 & 22,283 \\
  Density\footnotemark[1]  & $>$ 6.5 g/cm$^{3}$  & 4,602 \\
  PAL      & $<$ 17 cm           & 3,982 \\
  Dope site\footnotemark[2]  & yes                 & 1,825   \\ 
  LDA gap  & $>$ 3.0 eV          & 64   \\
  LDA gap  & $1.0<E_{g}<3.0$ eV          & 6   \\
  vbw      & $>$ 0.1 eV          & 62   \\
  cbw      & $>$ 0.2 eV          & 60    \\
  \hline
\footnotetext[1]{An upper limit of 13.0 g/cm$^{3}$ is applied for the density.} 
\footnotetext[2]{Compounds that pass this test must have a 3+ site or selected 2+ site. At least one of the following elements need to be present: La, Ce, Gd, Y, Lu, Sc, Be, Mg, Ca, Sr, Ba, Al, Ga, In, Tl, As, Sb, Bi. If Pr, Nd, Pm, Sm, Dy, Ho, Er, Tm, or an element with Z $>$ 83 is present the compound is excluded.} 
\end{tabular}
\caption{Results of the mining algorithm for host materials with Ce-activation when applied to Ref.1. A final list of 60 compounds is listed, which becomes a list of 70 compounds if cbw and cbw are ignored and the sulphur containing small band gap compounds are included.}\label{tab:cemining}

\end{center}
\end{table}

\newpage

\section{Appendix C: Suplementary information 3}

\begin{center}
\begin{longtable}{llllllll}
\caption{Semiconducting materials}\label{tab:semiminingres}\\


   \multicolumn{1}{l}{\textbf{Material}} &
   \multicolumn{1}{l}{\textbf{Spgrp}} &
   \multicolumn{1}{l}{\textbf{$\rho$}} &
   \multicolumn{1}{l}{\textbf{PAL}} &
   \multicolumn{1}{l}{\textbf{E$_{gap}$}} &
   \multicolumn{1}{l}{\textbf{Gap type}} &
   \multicolumn{1}{l}{\textbf{wbw/cbw}} &
   \multicolumn{1}{l}{\textbf{ICSD no.}} \\[0.5ex] 
   \\[-1.8ex]
\endfirsthead

\multicolumn{8}{c}{{\tablename} \thetable{} -- Continued} \\[0.5ex]
   \multicolumn{1}{l}{\textbf{Material}} &
   \multicolumn{1}{l}{\textbf{Spgrp}} &
   \multicolumn{1}{l}{\textbf{$\rho$}} &
   \multicolumn{1}{l}{\textbf{PAL}} &
   \multicolumn{1}{l}{\textbf{E$_{gap}$}} &
   \multicolumn{1}{l}{\textbf{Gap type}} &
   \multicolumn{1}{l}{\textbf{wbw/cbw}} &
   \multicolumn{1}{l}{\textbf{ICSD no.}} \\[0.5ex] 
  \\[-1.8ex]
\endhead

  \multicolumn{8}{l}{{Continued on Next Page\ldots}} \\
\endfoot

\endlastfoot
             AgHg$_{2}$O$_{4}$P &   55 & 8.2 & 2.5 & 1.35 &     direct & 0.46/1.81 &   2208\\
                    AgI$_{2}$Tl &  140 & 7.1 & 3.7 & 1.14 &     direct & 0.46/1.77 &  23159\\
                     AgLaOS     &  129 & 6.6 & 9.4 & 1.18 &     direct & 1.30/1.96 &  15530\\
              Ag$_{2}$HgO$_{2}$ &   96 & 9.3 & 2.8 & 0.51 &     direct & 0.43/1.35 & 280333\\
                      Ag$_{2}$S &   14 & 7.3 & 12.0 & 0.57 &     direct & 0.61/2.07 &  44507\\
              Ag$_{3}$LiO$_{2}$ &   72 & 7.1 & 11.9 & 0.65 &     direct & 0.64/1.31 &   4204\\
                    AlO$_{2}$Tl &  166 & 7.3 & 2.3 & 1.59 &     direct & 0.65/0.70 &  29010\\
                    AsLuO$_{4}$ &  141 & 6.9 & 5.0 & 3.40 &     direct & 0.84/1.92 &   2506\\
              As$_{2}$Eu$_{4}$O &  139 & 6.9 & 5.3 & 0.78 &     direct & 1.01/1.37 &   1222\\
                           AuBr &  138 & 8.2 & 2.5 & 1.42 &     direct & 1.30/1.01 & 200287\\
                           AuBr &  141 & 8.2 & 2.4 & 1.68 &     direct & 0.55/1.10 & 200286\\
                           AuCl &  141 & 7.8 & 2.2 & 1.30 &     direct & 0.84/1.12 &   6052\\
                            AuI &  138 & 8.3 & 2.5 & 1.42 &     direct & 1.10/0.72 &  24268\\
                          AuLiS &   70 & 7.0 & 2.5 & 1.34 &     direct & 0.87/1.37 & 280534\\
        Au$_{4}$S$_{3}$Tl$_{2}$ &   59 & 10.2 & 1.5 & 0.86 &     direct & 0.62/0.80 &  51235\\
                            BaO &  129 & 8.2 & 6.1 & 1.84 &     direct & 1.36/3.81 &  15301\\
                           BaSe &  221 & 6.6 & 9.8 & 1.08 &     direct & 3.48/6.67 &  52695\\
                           BiFO &  129 & 9.3 & 1.6 & 2.55 &     direct & 1.13/1.65 &  24096\\
                           BiIO &  129 & 9.7 & 1.9 & 0.70 &     direct & 1.49/2.78 &  29145\\
                    BiO$_{4}$Sb &   15 & 8.5 & 2.5 & 2.53 &     direct & 0.53/0.89 &  75901\\
              Bi$_{2}$O$_{6}$Te &   64 & 9.1 & 2.0 & 1.84 &     direct & 0.72/1.23 &   6239\\
 Bi$_{6}$Cu$_{2}$Pb$_{2}$S$_{12}$ &   26 & 7.0 & 2.3 & 0.45 &     direct & 0.65/0.82 &  95926\\
                          BrFPb &  129 & 7.7 & 2.4 & 2.49 &     direct & 1.14/1.60 &  30288\\
                           BrTl &  221 & 7.5 & 2.4 & 1.75 &     direct & 1.81/4.11 &  61532\\
                           BrTl &  225 & 6.6 & 2.7 & 1.92 &     direct & 2.20/3.00 &  61519\\
              Br$_{4}$STl$_{6}$ &  128 & 7.4 & 2.3 & 1.74 &     direct & 0.86/0.67 &  40521\\
             Br$_{6}$HgTl$_{4}$ &  128 & 7.0 & 2.7 & 1.81 &     direct & 0.60/0.82 &   9325\\
                     CO$_{3}$Pb &   62 & 6.6 & 2.5 & 2.97 &     direct & 0.46/0.90 &  36164\\
                    CaHgO$_{2}$ &  166 & 6.5 & 2.9 & 2.42 &     direct & 0.41/1.72 &  80717\\
                    CdHgO$_{2}$ &   12 & 9.5 & 2.3 & 0.60 &     direct & 0.61/2.52 &  74848\\
              CdI$_{6}$Tl$_{4}$ &  128 & 6.9 & 3.1 & 1.78 &     direct & 0.43/0.87 &  60756\\
                          ClFPb &  129 & 7.2 & 2.3 & 3.04 &     direct & 1.01/1.53 &  30287\\
      ClO$_{12}$P$_{3}$Pb$_{5}$ &  176 & 7.2 & 2.4 & 2.44 &     direct & 0.49/0.69 &  24238\\
                  ClO$_{2}$PbSb &   63 & 7.0 & 3.1 & 1.64 &     direct & 0.76/1.85 &  86229\\
        Cl$_{2}$Hg$_{7}$O$_{3}$ &   57 & 9.6 & 1.6 & 0.60 &     direct & 0.64/1.42 &  83225\\
              Cl$_{4}$STl$_{6}$ &  128 & 7.1 & 2.1 & 1.46 &     direct & 1.05/0.80 &  35289\\
              CrHg$_{5}$O$_{6}$ &   15 & 8.9 & 1.8 & 0.96 &     direct & 0.44/0.81 &  81605\\
                            CsI &  221 & 9.0 & 5.5 & 1.54 &     direct & 5.34/9.04 &  56524\\
                            CuI &  129 & 6.9 & 10.6 & 0.98 &     direct & 1.63/3.48 &  78268\\
              Eu$_{2}$O$_{4}$Si &   62 & 6.7 & 5.8 & 3.91 &     direct & 0.53/1.81 &   1510\\
                           FIPb &  129 & 7.4 & 2.6 & 1.50 &     direct & 1.10/1.44 & 279599\\
                           FInO &   70 & 6.6 & 13.1 & 1.62 &     direct & 0.77/4.03 &   2521\\
                            FTl &  139 & 8.4 & 1.7 & 1.82 &     direct & 2.87/5.95 &   9893\\
                            FTl &   28 & 9.0 & 1.6 & 1.37 &     direct & 1.12/2.14 &  16112\\
                            FTl &   69 & 8.5 & 1.7 & 1.73 &     direct & 2.94/6.06 &  30268\\
                      F$_{2}$Hg &  225 & 9.3 & 1.8 & 0.41 &     direct & 0.50/4.25 &  33614\\
              F$_{7}$SiTl$_{3}$ &  163 & 6.8 & 2.4 & 3.10 &     direct & 0.43/1.49 &  68021\\
             Gd$_{3}$InSe$_{6}$ &   58 & 7.2 & 7.4 & 0.67 &     direct & 0.65/1.08 & 280242\\
                      HfO$_{2}$ &  225 & 10.4 & 2.3 & 3.71 &     direct & 1.69/1.03 &  53033\\
                    HfO$_{3}$Pb &   55 & 10.2 & 1.7 & 2.27 &     direct & 0.43/1.15 &  52030\\
              HgI$_{6}$Tl$_{4}$ &  128 & 7.2 & 2.7 & 1.19 &     direct & 0.49/1.14 &  14018\\
                    HgO$_{3}$Ti &  161 & 8.7 & 2.4 & 1.25 &     direct & 0.57/1.01 &  19005\\
                     HgO$_{4}$W &   15 & 9.2 & 2.0 & 2.20 &     direct & 0.44/0.73 & 280911\\
              Hg$_{2}$O$_{3}$Se &   14 & 8.0 & 2.3 & 2.25 &     direct & 0.67/0.65 & 412302\\
         Hg$_{4}$N$_{2}$O$_{8}$ &   14 & 7.5 & 2.2 & 1.70 &     direct & 0.57/0.66 &  59156\\
        Hg$_{6}$O$_{7}$Si$_{2}$ &   12 & 9.1 & 1.8 & 1.56 &     direct & 0.47/1.01 &  69123\\
              ISe$_{2}$Tl$_{5}$ &  140 & 8.6 & 1.9 & 0.68 &     direct & 0.74/0.74 &  49524\\
                            ITl &  225 & 6.6 & 2.8 & 1.81 &     direct & 1.89/2.96 &  60491\\
               I$_{4}$STl$_{6}$ &  128 & 7.2 & 2.4 & 1.61 &     direct & 0.65/0.79 &  29265\\
              I$_{4}$SeTl$_{6}$ &  128 & 7.4 & 2.4 & 1.53 &     direct & 0.64/0.82 &  40520\\
                      O$_{2}$Sn &  136 & 6.9 & 11.8 & 0.52 &     direct & 1.27/5.15 &  39178\\
                      O$_{2}$Sn &   58 & 7.4 & 11.1 & 1.47 &     direct & 1.08/5.42 &  56675\\
                    O$_{3}$SbTl &  163 & 7.1 & 3.0 & 1.88 &     direct & 0.77/0.96 &   4123\\
                       O$_{3}$W &    7 & 7.4 & 3.1 & 1.50 &     direct & 0.42/1.64 &  84144\\
                O$_{4}$Pb$_{3}$ &  117 & 8.7 & 1.6 & 1.16 &     direct & 0.63/1.61 &  29094\\
                O$_{4}$Pb$_{3}$ &  135 & 8.9 & 1.6 & 0.64 &     direct & 0.73/1.51 &  22325\\


\end{longtable}
\end{center}


\begin{center}
\begin{longtable}{llllllll}
\caption{Cerium activated materials. 
}\label{tab:ceminingres}\\

   \multicolumn{1}{l}{\textbf{Material}} &
   \multicolumn{1}{l}{\textbf{Spgrp}} &
   \multicolumn{1}{l}{\textbf{$\rho$}} &
   \multicolumn{1}{l}{\textbf{PAL}} &
   \multicolumn{1}{l}{\textbf{E$_{gap}$}} &
   \multicolumn{1}{l}{\textbf{Gap type}} &
   \multicolumn{1}{l}{\textbf{wbw/cbw}} &
   \multicolumn{1}{l}{\textbf{ICSD no.}} \\[0.5ex] 
   \\[-1.8ex]
\endfirsthead

\multicolumn{8}{c}{{\tablename} \thetable{} -- Continued} \\[0.5ex]
   \multicolumn{1}{l}{\textbf{Material}} &
   \multicolumn{1}{l}{\textbf{Spgrp}} &
   \multicolumn{1}{l}{\textbf{$\rho$}} &
   \multicolumn{1}{l}{\textbf{PAL}} &
   \multicolumn{1}{l}{\textbf{E$_{gap}$}} &
   \multicolumn{1}{l}{\textbf{Gap type}} &
   \multicolumn{1}{l}{\textbf{wbw/cbw}} &
   \multicolumn{1}{l}{\textbf{ICSD no.}} \\[0.5ex] 
  \\[-1.8ex]
\endhead

  \multicolumn{8}{l}{{Continued on Next Page\ldots}} \\
\endfoot

\endlastfoot

                    AlLaO$_{3}$ &  167 & 6.5 & 9.9 & 3.68 &  in-direct & 0.29/0.66 &  90536\\
                    AlO$_{3}$Tb &   62 & 7.5 & 5.3 & 6.38 &     direct & 0.32/2.38 &  84422\\
      Al$_{2}$Gd$_{2}$O$_{7}$Sr &  139 & 6.9 & 7.2 & 5.90 &     direct & 0.31/3.27 &  33580\\
       Al$_{3}$F$_{19}$Pb$_{5}$ &  108 & 6.8 & 2.7 & 4.56 &  in-direct & 0.24/0.26 & 203224\\
       Al$_{3}$F$_{19}$Pb$_{5}$ &  140 & 6.7 & 2.8 & 4.18 &  in-direct & 0.21/0.41 &  96597\\
       Al$_{3}$F$_{19}$Pb$_{5}$ &   87 & 6.7 & 2.8 & 4.37 &  in-direct & 0.19/0.33 &  80105\\
                    AsBiO$_{4}$ &   88 & 7.7 & 2.6 & 3.03 &  in-direct & 0.25/0.70 &  30636\\
                    AsLuO$_{4}$ &  141 & 6.9 & 5.0 & 3.40 &     direct & 0.84/1.92 &   2506\\
                   BGaO$_{4}$Pb &   62 & 6.9 & 3.2 & 3.28 &  in-direct & 0.58/0.90 & 279600\\
                     BLuO$_{3}$ &  167 & 6.9 & 3.9 & 5.27 &  in-direct & 0.34/0.84 &  16525\\
            BaBeLa$_{2}$O$_{5}$ &   14 & 6.6 & 7.9 & 3.78 &  in-direct & 0.36/0.16 &  65292\\
                      BaF$_{2}$ &   62 & 6.7 & 8.6 & 5.52 &  in-direct & 0.92/2.43 &  41651\\
                    BaO$_{3}$Tb &   62 & 7.3 & 5.3 & 4.45 &     direct & 1.02/1.96 &  86736\\
              BaO$_{4}$Tb$_{2}$ &   62 & 7.8 & 4.5 & 3.89 &  in-direct & 0.15/1.83 &  78661\\
            BaO$_{5}$Tb$_{2}$Zn &   62 & 7.8 & 5.1 & 3.72 &     direct & 0.38/1.15 &  69721\\
  Ba$_{2}$Ce$_{0.75}$O$_{6}$Sb  &  139 & 6.5 & 8.6 & 4.00 &  in-direct & 0.49/0.81 &  72522\\
            Ba$_{2}$EuO$_{6}$Sb &  225 & 7.0 & 7.0 & 4.12 &  in-direct & 0.30/1.26 &  38330\\
            Ba$_{2}$GdO$_{6}$Sb &  225 & 7.1 & 6.8 & 3.49 &     direct & 0.37/1.56 &  38331\\
            Ba$_{2}$O$_{6}$SbTb &  225 & 7.2 & 6.5 & 4.18 &     direct & 0.32/1.27 &  38332\\
            Ba$_{2}$O$_{6}$SbYb &  225 & 7.5 & 5.4 & 4.19 &     direct & 0.35/1.44 &  38336\\
            Ba$_{2}$O$_{6}$TaYb &  225 & 8.2 & 3.7 & 3.49 &  in-direct & 0.38/0.83 &  91001\\
             Ba$_{2}$O$_{6}$WZn &  225 & 7.7 & 5.0 & 3.37 &  in-direct & 0.31/0.56 &  24983\\
      Ba$_{3}$NiO$_{9}$Ru$_{2}$ &  194 & 6.8 & 10.6 & 3.97 &     direct & 0.09/1.73 &  50832\\
       Ba$_{4}$O$_{10}$Ru$_{3}$ &   64 & 6.7 & 9.9 & 4.72 &  in-direct & 0.12/0.74 &  90902\\
                    BeF$_{4}$Pb &   62 & 6.8 & 2.7 & 4.34 &  in-direct & 0.19/0.53 &  24568\\
                      BiF$_{3}$ &   62 & 7.9 & 2.0 & 3.81 &  in-direct & 0.58/0.68 &   1269\\
       Bi$_{4}$Ge$_{3}$O$_{12}$ &  220 & 7.1 & 2.6 & 3.26 &  in-direct & 0.22/0.32 &  39231\\
       Bi$_{4}$O$_{12}$Si$_{3}$ &  220 & 6.8 & 2.4 & 3.58 &  in-direct & 0.37/0.24 &  84519\\
                     BiO$_{4}$V &   15 & 7.0 & 2.8 & 3.38 &  in-direct & 0.39/0.51 &  31549\\
                          BrGdO &  129 & 6.8 & 6.4 & 4.16 &     direct & 1.08/0.99 &  41071\\
              CaLu$_{2}$O$_{4}$ &   62 & 8.1 & 3.2 & 3.49 &  in-direct & 0.42/1.23 &  15125\\
             CaO$_{11}$Ta$_{4}$ &  182 & 7.6 & 3.3 & 3.08 &  in-direct & 0.18/0.38 &   1854\\
              CaO$_{4}$Yb$_{2}$ &   62 & 8.0 & 3.4 & 4.38 &     direct & 0.22/1.65 &  27312\\
              CaO$_{6}$Ta$_{2}$ &   62 & 7.4 & 3.5 & 3.15 &  in-direct & 0.17/0.80 &  24091\\
                      CeO$_{2}$ &  225 & 7.2 & 6.8 & 5.62 &  in-direct & 0.48/1.38 &  28753\\
                Ce$_{2}$O$_{3}$ &  164 & 6.5 & 7.1 & 3.61 &  in-direct & 0.43/1.36 &  96197\\
                          ClGdO &  129 & 6.7 & 5.8 & 5.13 &     direct & 0.72/1.13 &  59232\\
                      F$_{3}$La &  139 & 7.0 & 8.5 & 5.40 &  in-direct & 0.88/0.40 &  96133\\
              F$_{6}$SnTl$_{2}$ &  164 & 6.8 & 2.9 & 4.03 &     direct & 0.78/1.29 & 410801\\
              F$_{7}$SiTl$_{3}$ &  163 & 6.8 & 2.4 & 3.10 &     direct & 0.43/1.49 &  68021\\
                    GaLaO$_{3}$ &  161 & 7.0 & 10.4 & 3.27 &  in-direct & 0.18/0.78 &  51039\\
                    GaLaO$_{3}$ &  167 & 6.9 & 10.5 & 3.01 &  in-direct & 0.21/1.02 &  51286\\
                    GaLaO$_{3}$ &   62 & 7.2 & 10.1 & 3.24 &     direct & 0.51/0.56 &  79662\\
              Gd$_{2}$GeO$_{5}$ &   14 & 7.1 & 5.9 & 3.76 &     direct & 0.24/1.16 &  61372\\
                Gd$_{2}$O$_{3}$ &  206 & 7.6 & 4.4 & 3.20 &  in-direct & 0.23/1.50 &  40473\\
              Gd$_{2}$O$_{5}$Si &   14 & 6.8 & 5.7 & 4.71 &  in-direct & 0.17/0.78 &  27728\\
       Gd$_{3}$O$_{12}$Sb$_{5}$ &  217 & 6.6 & 7.3 & 3.16 &  in-direct & 0.13/0.17 &  65147\\
        Ge$_{2}$Lu$_{2}$O$_{7}$ &   92 & 7.4 & 4.6 & 3.55 &  in-direct & 0.12/1.53 &  39929\\
       Ge$_{4}$Lu$_{6}$O$_{17}$ &   13 & 7.4 & 4.1 & 3.85 &  in-direct & 0.16/1.38 &  39790\\
                    InO$_{4}$Ta &   13 & 8.3 & 3.9 & 3.54 &  in-direct & 0.24/0.60 &  72569\\
               KO$_{8}$W$_{2}$Y &   15 & 6.6 & 4.6 & 3.32 &     direct & 0.18/0.22 &  90378\\
                    LaO$_{3}$Yb &   33 & 8.2 & 3.8 & 4.30 &     direct & 0.25/0.35 &  30399\\
            La$_{2}$LiO$_{6}$Sb &   14 & 6.5 & 9.0 & 3.82 &  in-direct & 0.21/0.74 &  72202\\
                La$_{2}$O$_{3}$ &  150 & 6.6 & 7.5 & 3.16 &  in-direct & 0.75/1.13 &  56166\\
                La$_{2}$O$_{3}$ &  164 & 6.5 & 7.6 & 3.74 &  in-direct & 0.38/0.83 &  96196\\
                    LuO$_{2}$Rb &  166 & 7.6 & 4.2 & 3.48 &  in-direct & 0.53/1.83 &  15164\\
                     LuO$_{4}$P &  141 & 6.5 & 4.7 & 5.54 &     direct & 0.89/1.57 &  79761\\
                Lu$_{2}$O$_{3}$ &  206 & 9.4 & 2.4 & 3.77 &  in-direct & 0.28/1.21 &  40471\\
              Lu$_{2}$O$_{5}$Si &   14 & 7.9 & 3.3 & 4.65 &  in-direct & 0.24/1.12 &  89624\\
             O$_{11}$SrTa$_{4}$ &  182 & 7.8 & 3.3 & 3.07 &  in-direct & 0.19/0.37 &  79704\\
       O$_{12}$Sb$_{5}$Yb$_{3}$ &  217 & 7.0 & 5.4 & 3.15 &  in-direct & 0.09/0.25 &  20945\\
               O$_{4}$STl$_{2}$ &   62 & 6.8 & 2.4 & 3.68 &  in-direct & 0.31/0.61 &  27440\\
              O$_{4}$SeTl$_{2}$ &   62 & 7.0 & 2.5 & 3.46 &  in-direct & 0.22/0.58 &  73411\\
              O$_{4}$SrYb$_{2}$ &   62 & 8.4 & 3.5 & 4.78 &     direct & 0.19/1.65 &  15123\\
                     AgLaOS     &  129 & 6.6 & 9.4 & 1.18 &     direct & 1.30/1.96 &  15530\\
            AsBrHg$_{3}$S$_{4}$ &  186 & 6.6 & 3.1 & 1.28 &  in-direct & 0.56/1.34 & 280330\\
                          BiBrS &   62 & 6.5 & 2.8 & 1.47 &  in-direct & 0.61/1.32 &  31389\\
                           BiIS &   62 & 6.8 & 2.8 & 1.16 &  in-direct & 0.62/1.34 &  23631\\
                Bi$_{2}$S$_{3}$ &   62 & 6.8 & 2.2 & 1.04 &  in-direct & 0.59/0.97 & 201066\\
                           FLuS &  166 & 6.7 & 3.9 & 2.40 &  in-direct & 1.24/1.35 &  89549\\

\end{longtable}
\end{center}


\end{widetext}

\begin{thebibliography}{10}

\bibitem{website}
\url{http://gurka.fysik.uu.se/esp/}.

\bibitem{cohen00}
M.~Cohen.
\newblock ''The theory of real materials''
\newblock {\em Annual Review of Material Science}, 30:1--26, 2000.

\bibitem{Olson2000a}
G.~B. Olson.
\newblock {''Materials by Designing a New Material World''}
\newblock {\em Science}, 288(5468):993--998, 2000.

\bibitem{andersson:037205}
G.~Andersson, T.~Burkert, P.~Warnicke, M.~Bj\"{o}rck, B.~Sanyal, C.~Chacon,
  C.~Zlotea, L.~Nordstr\"{o}m, P.~Nordblad, and O.~Eriksson.
\newblock ''Perpendicular magnetocrystalline anisotropy in tetragonally distorted
  Fe-Co alloys''
\newblock {\em Physical Review Letters}, 96(3):037205, 2006.

\bibitem{burkert:027203}
T.~Burkert, L.~Nordstr\"{o}m, O.~Eriksson, and O.~Heinonen.
\newblock ''Giant magnetic anisotropy in tetragonal FeCo alloys''
\newblock {\em Physical Review Letters}, 93(2):027203, 2004.

\bibitem{Hart:2005lr}
G.~L.~W. Hart, V.~Blum, M.~J. Walorski, and A.~Zunger.
\newblock ''Evolutionary approach for determining first-principles hamiltonians''
\newblock {\em Nature Materials}, 4(5):391--394, 2005.

\bibitem{ceder}
C.~C. Fischer, K.~J. Tibbetts, D.~Morgan, and G.~Ceder.
\newblock ''Predicting crystal structure by merging data mining with quantum mechanics''
\newblock {\em Nature Materials}, 5:641--646, 2006.

\bibitem{icsd}
ICSD Inorganic Crystal Structure Database, FIZ Karlsruhe,
  \url{http://www.fiz-karlsruhe.de/icsd.html}.

\bibitem{wills}
J.~M. Wills, O.~Eriksson, M.~Alouani, and D.~L. Price.
\newblock {\em ''Electronic Structure and Physical Properties of solids: The uses
  of the LMTO method''}
\newblock Springer Verlag, Berlin, 2000.

\bibitem{db1}
\url{http://databases.fysik.dtu.dk/}.

\bibitem{db2}
\url{http://caldb.nims.go.jp/}

\bibitem{db3}
\url{http://cst-www.nrl.navy.mil/}

\bibitem{db4}
\url{http://ptp.ipap.jp/link?PTPS/138/755/}



\bibitem{bradleycracknell}
C.~J. Bradley, and A.~P. Cracknell
\newblock {\em ''The Mathematical Theory of Symmetry in Solids: Represetation theory for point groups and space groups''}
\newblock Clarendon press, Oxford, 1972.

%
%
\bibitem{lehman}
W.~Lehmann.
\newblock ''Edge emission of n-type conducting ZnO and CdS''
\newblock {\em Solid-State Electronics}, 9:1107--1110, 1966.

\bibitem{knoll}
G.~F. Knoll.
\newblock {\em Radiation Detection and Measuremnet}.
\newblock John Wiley and Sons, 2000.

\bibitem{atalla}
M.~M. Atalla.
\newblock {\em Scientific Rept. no. 8}, page AD0447260, 1964.

\bibitem{crchandbook}
D.~R. Lide, editor.
\newblock {\em CRC Handbook of Chemistry and Physics}.
\newblock Taylor and Francis, Boca Raton, FL. Internet version, 88th edition
  edition, 2007.

\bibitem{yen}
W.~Yen, M.~Raukas, S.~Basun, W.~van Schaik, and U.~Happek.
\newblock ''Optical and photoconductive properties of cerium-doped crystalline
  solids''
\newblock {\em Journal of Luminescence}, 69:287--294, 1996.

\bibitem{dorenbos}
P.~Dorenbos.
\newblock ''The 5d level positions of the trivalent lanthanides in inorganic
  compounds''
\newblock {\em Journal of Luminescence}, 91:155--176, 2000.

\end{thebibliography}
\end{document}